\documentclass[a4paper,floatsintext,man,natbib]{apa6}
\usepackage[english]{babel}
\usepackage[utf8]{inputenc}
\usepackage[pdftex, pdftitle={Article}, pdfauthor={Author}]{hyperref} 
\usepackage{amsthm,amsfonts, amssymb, amsmath}
\usepackage{physics}
\usepackage{booktabs}
\usepackage{xcolor}
\usepackage{graphicx}
\usepackage{adjustbox}
\usepackage{placeins}
\usepackage[T1]{fontenc}
\usepackage{lipsum}
\usepackage{subfigure}
\usepackage{csquotes}
\usepackage{natbib}
\usepackage{multirow}
\usepackage{tikz}
\setcitestyle{authoryear,open={(},close={)}}

\theoremstyle{definition}

\newcommand{\key}{\textbf}

\title{An information-theoretic account of semantic interference in word production}
\shorttitle{Information-theoretic account of interference}
\affiliation{University of California, Irvine}

\abstract{
I present a computational-level model of semantic interference effects in word production. Word production is cast as a rate--distortion problem where an agent selects words to minimize a measure of cost while also minimizing the resources used to compute the output word based on perceptual input and behavioral goals. I show that similarity-based interference among words arises naturally in this setup, and I present a series of simulations showing that the model captures some of the key empirical patterns observed in Stroop and Picture--Word Interference paradigms. I argue that the rate--distortion account of interference provides a high-level formalization of computational principles that are instantiated more mechanistically in existing models.
}

\author{Richard Futrell}

\begin{document}

\maketitle

In cognitive science and related fields, \key{bounded rationality} is the idea that our cognitive systems are designed to take actions that are approximately optimal, given that only limited computational resources are available for calculating the optimal action \citep{simon1955behavioral,simon1972theories,kahneman2003maps,howes2009rational,lewis2014computational,gershman2015computational,lieder2020resource}. The idea is appealing because it maintains the mathematical precision of theories based on rationality, while avoiding the paradoxes and empirical shortcomings that come from claiming that human beings act in ways that are entirely rational. There has been recent interest in formalizing bounded rationality within the mathematical framework of rate--distortion theory \citep{berger1971rate,cover2006elements} with applications to cognitive science \citep{sims2016rate,sims2018efficient,zaslavsky2018efficient,gershman2020origin,zaslavsky2020rate}. 



In this paper, I apply rate--distortion theory to derive a model of word production. The main contribution of this paper is to show that rate--distortion theory generically predicts \key{semantic interference effects} when a subject is trying to produce a target word in the presence of a semantically related distractor. For example, the Stroop task famously exhibits interference \citep{stroop1935studies}: given a stimulus such as the word \textbf{\textcolor{red}{BLUE}} printed in red ink, and an instruction to name the color of the ink, it is hard to produce `red' because of interference from the similar word `blue'. A similar kind of interference is present in the Picture--Word Interference task, where a drawing must be named in the presence of a superimposed distractor word \citep{lupker1979semantic,starreveld2017picture}. I show that rate--distortion theory predicts a number of key phenomena observed in such tasks.


\section{Background: Rate--distortion theory of control}
\label{sec:rdcc}

\subsection{Bounded rationality}

Ultimately, our cognitive systems implement an \key{action policy}: a function from sensory inputs to motor outputs. For example, an animal might see another animal and decide among a large set of possible actions, including attacking, approaching, ambushing, fleeing, etc. In general, we can conceive of an action policy as a stochastic function mapping states $S$ (including perceptual, physiological, and memory information) to probability distributions on motor actions $A$:
\begin{equation*}
    q : S \rightarrow A.
\end{equation*}
We can also think of the policy as a \emph{probability distribution} on actions given states, where $q(a|s)$ denotes the probability of taking action $a$ in state $s$.


A \key{boundedly rational action policy} is a policy that chooses an action to maximize some measure of reward, or equivalently, to minimize the cost of the \emph{consequences} of taking a certain action in the world, subject to a constraint on the computational resources used in finding and implementing this action. These resources include factors such as time---in many circumstances, it may be more important to act quickly than to take the time to compute the best action---as well as physiological resources such as the energy required to perform computations.
Formally, letting $D(s,a)$ represent the \key{action cost} or the cost of the consequences of taking action $a$ in state $s$, and letting $C(s,a)$ denote the \key{computation cost} required to compute the action $a$ given state $s$, then the overall cost for a policy $q$ is
\begin{equation}
    \label{eq:control-objective}
    \mathcal{L}(q) = \Big \langle D(s,a) + \frac{1}{\gamma} C(s,a) \Big \rangle,
\end{equation}
where $\big \langle \cdot \big \rangle$ denotes an average over the joint probability distribution on states and actions given those states $p(s)q(a|s)$, and $\frac{1}{\gamma}$ is a scalar value which indicates how much a unit of computation cost $C$ should be weighed against a unit of action cost $D$.
The scalar $\gamma$ can also be viewed as a parameter giving the amount of resources available for computation: high $\gamma$ means that the agent is willing to perform a lot of computation in order to minimize the action cost $D$.

The expression $\mathcal{L}(q)$ in (\ref{eq:control-objective}) is called the \key{control objective}, and a bounded optimal action policy is derived by minimizing it:
\begin{equation*}
    q_{\text{bounded rational}} = \arg \min_q \mathcal{L}(q),
\end{equation*}
where the minimization is over the set of all possible policies. 

Without further specifications, the theory of bounded rationality goes no farther than the formalization above. Given a set of cost functions, the bounded rational action policy is derived as the solution to a multi-objective minimization problem involving those cost functions. The theory only makes precise predictions when the cost functions and their relative weights are further specified. Below, we will see how we can do this in a principled way using tools from information theory.

\subsection{Rate--distortion theory}
\label{sec:rd}

Rate-distortion theory is the mathematical theory of lossy communication and compression, a subfield of information theory. It provides mathematical tools to answer questions like: if I want to transmit a picture of a zebra to you, and I do not have the capacity to send it to you perfectly, how can I encode the image such that your received picture looks approximately like what I sent? This problem involves two constraints: (1) my capacity to transmit information (called \key{rate}), and (2) a measure of how much your received picture differs from my picture (this measure is called \key{distortion}). Rate--distortion theory describes the problem of finding a data encoding which minimizes the distortion subject to a constraint on the rate. 

The link between rate--distortion theory and bounded rational action policies was not immediately clear, although the original paper on rate--distortion theory did note a connection with control theory in passing \citep{shannon1959coding}. The key insight that has enabled researchers to link these two theories is that rate--distortion theory can be applied to constrain the perception--action loop. The idea is to treat an action policy as a communication channel from sensory input to motor output. Then the action cost $D$ in Eq.~\ref{eq:control-objective} is the distortion, and the computation cost $C$ in Eq.~\ref{eq:control-objective} is the rate. This connection was introduced first in the economics literature by \citet{sims2003implications,sims2005rational,sims2010rational}, then picked up in the robotics, cybernetics, machine learning, and psychology literature \citep[][among others]{vandijk2009hierarchical,tishby2011information,rubin2012trading,ortega2013thermodynamics,genewein2015bounded,sims2016rate,sims2018efficient,gershman2020rationally}. 

In the \key{rate--distortion theory of control} (RDC), a boundedly rational action policy is derived by minimizing the following control objective:
\begin{equation}
    \label{eq:rdcc}
    \mathcal{L}(q) = \Big \langle D(s, a) \Big \rangle + \frac{1}{\gamma} I[S:A],
\end{equation}
where $D(s,a)$ is the distortion or action cost for taking action $a$ in state $s$, and $I[S:A]$ denotes the \key{mutual information} between the random variables $S$ representing the state and $A$ representing the action policy:
\begin{equation*}
    I[S:A] = \Bigg \langle \log \frac{q(a|s)}{q(a)} \Bigg \rangle,
\end{equation*}
where the probability $q(a)$ is the marginal probability of taking action $a$ under the policy $q$, averaging over all states:
\begin{equation}
    \label{eq:marginal}
    q(a) = \sum_s p(s) q(a|s).
\end{equation}

The substantive claim of the RDC is that computation costs should be modeled as the mutual information between states and actions $I[S:A]$. This quantity can be interpreted as the amount of information that must be extracted from $S$ in order to specify $A$ \citep{sims2003implications}, or as the information throughput of a controller implementing the policy $q(a|s)$ \citep{fan2014information}. I will argue below that this is a natural measure of computation cost, and that it subsumes many other measures.

I summarize four converging motivations for the use of the mutual information between states and actions $I[S:A]$ (and related measures such as relative entropy) as a measure of computation cost. I provide pointers into the literature for the full forms of these arguments. See also \citet[][\S 4]{zenon2019information} for a comprehensive discussion and review.
\begin{enumerate}

    \item \textbf{Computation time.} The mutual information reflects the \emph{search time} taken to find the action $A$ given state $S$ by a rejection sampling algorithm. When $I[S:A]$ is lower, the correct action can be found using fewer samples from $q(a)$. This argument is presented in more detail in \citet{braun2014information}. This perspective relates mutual information to the time taken to accumulate evidence for an action in sequential sampling models \citep{ratcliff2004comparison}.
    
    \item \textbf{Algorithmic complexity.} The mutual information reflects how many bits of information an agent must store to remember the policy, or how many bits of information an agent needs to see to learn the policy. This argument is presented in a PAC-Bayes framework by \citet{rubin2012trading}, who also show that action policies with a mutual information penalty are generically simpler and less prone to overfitting to their environment.

    \item \textbf{Free energy.} The RDC objective in Eq.~\ref{eq:control-objective} is technically a \key{free energy} functional, bringing the theory in line with neuroscientific theories of brain function formulated in terms of minimizing free energy \citep{friston2010free,lynn2020abstract}.
    
    \item \textbf{Congruence with empirically-observed laws of behavior.} A number of RDC theories have held that the time taken to initiate an action should be proportional to the amount of control information required to specify that action \citep{fan2014information}. Under this assumption, we can derive well-validated empirical laws of behavior as special cases. For example, Hick's Law is the observation that the time taken to decide among a set of actions $A$ is proportional to $\log |A|$ \citep{hick1952rate,hyman1953stimulus}. The RDC computation cost $I[S:A]$ reduces to $\log |A|$, yielding Hick's Law, in the case where (1) an agent is deciding among a set of actions $A$, (2) the default policy $q(a)$ is uninformative about which action to take, and (3) the policy $q(a|s)$ specifies the desired action deterministically. 
    Other generalizations such as Fitt's Law can be derived in similar ways \citep{zenon2019information}.
    \end{enumerate}
In summary, there is a convergence among a number of previous intuitive notions of computation cost, all of which point towards $I[S:A]$ as a reasonable measure. In addition to these theoretical arguments, a growing neuroscience literature has linked information measures such as $I[S:A]$ to brain activity in the prefrontal cortex \citep{koechlin2007information,fan2014information}.

The form of the RDC objective in Eq.~\ref{eq:rdcc} is only the simplest member of a family of possible control objectives. In reality, a cognitive agent must integrate information from many different inputs and produce motor output on many different actuators. Each input and each motor output can be associated with its own channel, with its own information-based penalty. Multiple input channels can be modeled by adding further weighted mutual information terms to Eq.~\ref{eq:rdcc}; such RDC policies have been studied by \citet{vandijk2011grounding,vandijk2013informational,genewein2015bounded}, and can be situated within the framework of Coherent Infomax \citep{kay2011coherent}. 

In fact, we will see that our model of Pi interference requires at least two input channels: a top-down goal signal and a bottom-up perceptual signal. In that setting, interference effect arises from a penalty on \emph{synergistic} information among three variables: the top-down goal signal, the bottom-up perceptual signal, and the output action.

\subsection{Solutions to the RDC Objective}
\label{sec:rd-solutions}

The policies admitted under the rate--distortion theory of control have a common mathematical form. The minima of Eq.~\ref{eq:rdcc} obey the following equations, derived by \citet{blahut1972computation}:\footnote{Briefly, this result is derived as follows. We take the derivative of $\mathcal{L}(q)$ with respect to the probability $q(a|s)$ of each action $a$ in a state $s$, including a Lagrange multiplier $\lambda$ that will be used to normalize the probability distribution. Then we set the result to zero, and solve for $q(a|s)$:
\begin{align}
    \nonumber
    \frac{\partial}{\partial q(a|s)} \big ( \mathcal{L} &+ \lambda \sum_a q(a|s) \big ) = 0\\
    \nonumber
    D(s, a) &+ \frac{1}{\gamma}\log \frac{q(a|s)}{q(a)} + \lambda^\prime = 0 \\
    \nonumber
    \log q(a|s) &= \log q(a) - \gamma D(s, a) - \lambda^{\prime\prime} \\
    \nonumber
    q(a|s) &= \frac{1}{Z} q(a) e^{-\gamma D(s, a)},
\end{align}
where $\lambda^\prime = \lambda + \frac{1}{\gamma}$, and $\lambda^{\prime\prime} = \gamma \lambda^\prime = \log Z$.}
\begin{align}
    \label{eq:equilibrium}
    q(a|s) &= \frac{1}{Z(s)} q(a) \exp{- \gamma D(s, a)} \\
    \nonumber
    q(a) &= \sum_s p(s) q(a|s) \\
    \nonumber
    Z(s) &= \sum_a q(a) \exp{- \gamma D(s, a)}.
\end{align}
Note that the equations in Eq.~\ref{eq:rdcc} do not specify a policy uniquely. The problem is that an expansion of the definition of $q(a|s)$ will require an expansion of the definition of $q(a)$, which itself contains $q(a|s)$ recursively. The equations are called self-consistent, meaning that any $q(a|s)$, $q(a)$, and $Z(s)$ constitute a minimum of the control objective as long as they satisfy the three equations simultaneously. In general, multiple solutions exist. A numerical solution to the equations can be found by starting with a random value of $q(a|s)$, then evaluating the equations iteratively until a fixed point is reached.

\subsection{Link to behavioral measures}

The RDC describes the derivation of boundedly rational action policies, but does not immediately make predictions about the timing of these actions nor other behavioral and neural dependent measures that are commonly deployed in the study of cognitive control and language production. A linking hypothesis is required from the mathematical policy $q(a|s)$ to predictions about dependent measures such as reaction time, the usual measure of difficulty in word production studies.

There are a number of perspectives in the psychological literature on the relationship between reaction times and information-theoretic measures of complexity \citep{laming1968information,laming2003human,luce2003whatever,fan2014information,zenon2019information,lynn2020abstract}. The simplest possible hypothesis is that the time required to initiate an action is linearly proportional to the amount of computation that needs to be done to select the action. I adopt this simple view here. Furthermore, the computation required to select an action breaks into multiple parts, which I call computation cost and decision cost:
\begin{enumerate}
    \item \textbf{Computation cost}. The computation required to produce the action policy $q(a|s)$. This is equal to the cost term in the policy objective $\mathcal{L}$ that generates $q(a|s)$. For example, given the control objective in Eq.~\ref{eq:rdcc}, the average computation cost is the mutual information $I[S:A] = \Big \langle \log \frac{q(a|s)}{q(a)} \Big \rangle$. For a particular action $a$ in state $s$, the cost is the pointwise mutual information $\log \frac{q(a|s)}{q(a)}$. This notion of computation cost combines \citet{zenon2019information}'s notions of `perceptual cost' and `automatic cost'. For human behavioral work relating this notion of computation cost to computation time, see \citet{ortega2016human} and \citet{schach2018quantifying}.
    \item \textbf{Decision cost}. A policy $q(a|s)$ is a probability distribution on actions, but in any given state, an agent must take a single action alone. Decision cost is the cost associated with selecting a single action $a^*$ from a distribution $q(a|s)$; it respresents a decision that still needs to be made (perhaps randomly) after calculation of the policy. I take decision cost to be equal to the KL divergence from $q(a|s)$ to a delta distribution specifying a single action $a^*$:
    \begin{align}
        \nonumber
        D_{\text{KL}}[\delta_{aa^*} || q(a|s)] &= \Bigg \langle \log \frac{\delta_{aa^*}}{q(a|s)} \Bigg \rangle \\
        &= -\log q(a^*|s),
    \end{align}
    where $\delta_{aa^*}$ is a Kronecker delta function (equal to 1 when $a = a^*$ and $0$ otherwise).
    This cost is identified with cognitive control by \citet{fan2014information}, who conceptualizes cognitive control as the means by which uncertainty is reduced, at a rate of $\sim 20$ ms per bit. In this case, the policy $q(a|s)$ represents the uncertainty to be reduced.
\end{enumerate}

It stands to reason that both computation cost and decision cost make contributions to dependent measures such as reaction time, although perhaps not according to a simple function. In this work I will present computation and decision cost in terms of bits of information, and where appropriate I will discuss their possible translation into psychological dependent measures.






%

To sum up this section, I have presented the rate--distortion theory of control (RDC) as a model of boundedly rational action. Below, I will present a new application of this model to model human word production, which exhibits a property of the model which has not previously been explored. In particular, I will show that similarity-based interference effects, which are common in word production as well as other aspects of cognition, are a generic prediction of RDC models.

\section{Interference in the Rate--Distortion Theory of Control}
\label{sec:interference}

In this section I will demonstrate the basic mechanism by which RDC predicts similarity-based interference effects. 

\subsection{The empirical phenomena}

The term \key{similarity-based interference} encompasses a large number of phenomena in human perception, action, and memory. It refers to the idea that percepts, actions, or memories are confused for each other when they are `similar' according to some metric \citep{shepard1987toward}, that is, when they share features or associated cues. Furthermore,  there may be increased latency in identifying a percept, retrieving information from memory \citep{jaeger2017similaritybased}, or initiating in action \citep{stroop1935studies} when a `similar' distractor is also available in some sense.  Capturing similarity-based interference is a key goal of cognitive models, including those based on cue-based retrieval, spreading activation, and production rules \citep{watkins1975buildup,ratcliff1978theory,anderson1998atomic,roelofs2003goal}.

\subsection{RDC account}

Similarity-based interference arises generically in RDC models because the action cost $D(s,a)$ naturally defines a similarity metric among actions \citep{sims2018efficient}. The function $D(s,a)$ gives the cost of taking action $a$ in state $s$. Two actions are similar when they have similar cost, that is, when there is low cost for failing to distinguish them. Accordingly, we can define a distance metric between two actions. In state $s$, let $a_s$ be the action with minimal cost, and $a_d$ be any other action. The state-dependent distance metric among actions can be defined as a function
\begin{equation}
    \label{eq:cost-diff}
    d(a_s, a_d) = D(s, a_d) - D(s, a_s).
\end{equation} 
This distance metric\footnote{The function $d(a_s, a_d)$ is technically a pre-metric. It satisfies $d(a,a)=0$ for all actions $a$, and it is always non-negative. It is non-negative because $a_s$ is defined as the action with minimal cost in state $s$. The function is only a pre-metric, not a full metric, because it is not generally symmetrical. That is, $d(a_s, a_d) \neq d(a_d, a_s)$ in general.} will play the role of the distortion metric in rate--distortion theory.


Now that we have a distance metric among actions, we can see that interference effects arise even in the simplest formulation of the RDC. Suppose the control system is attempting to solve the following problem: in a state $s$ (for example, seeing a picture of an apple), there is a single unique target action $a_s$ corresponding to that state (for example, saying the word ``apple''). The agent is attempting to generate the right target action in state $s$. In this setting, RDC predicts generally that the probability that any two actions (e.g., words) $a_s$ and $a_d$ are confused will increase as the distance $d(a_s, a_d)$ between them decreases---thus predicting similarity-based interference among competitors in word production.

\begin{figure}
    \centering
        \includegraphics[width=.3\textwidth]{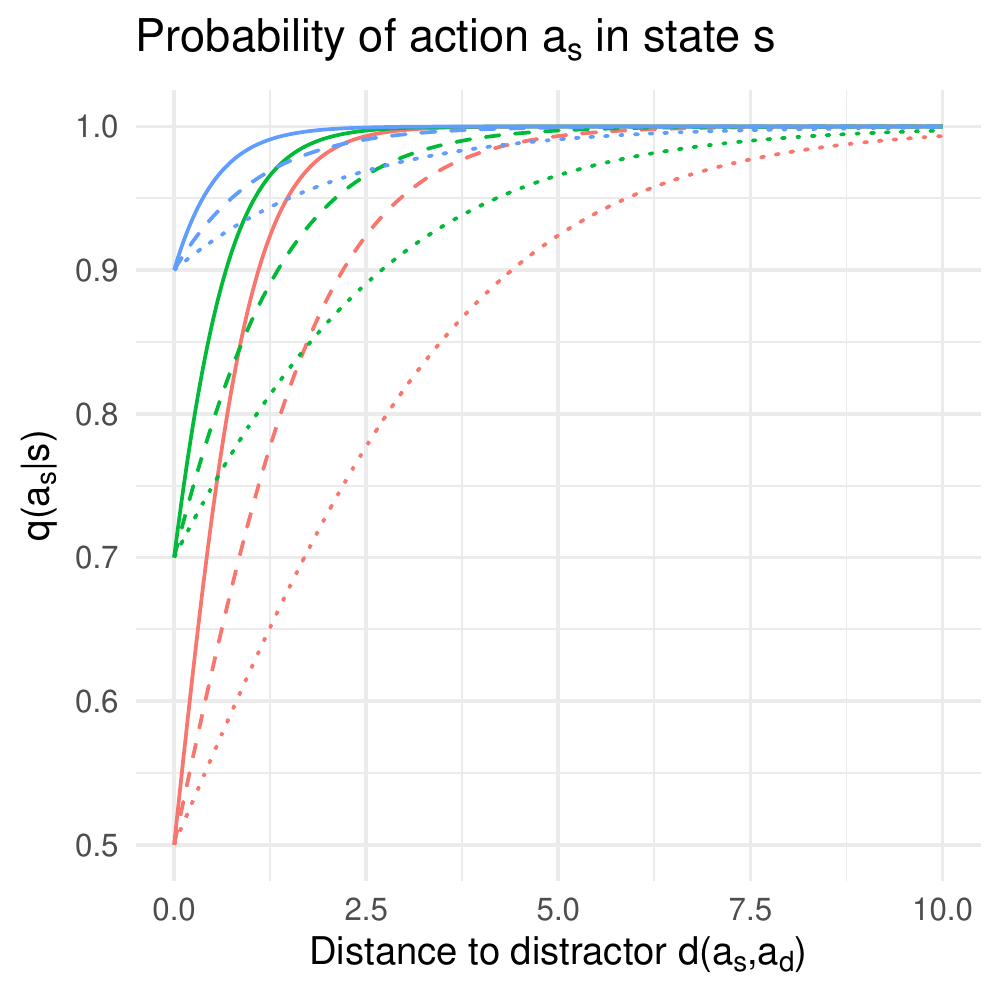}
        \includegraphics[width=.3\textwidth]{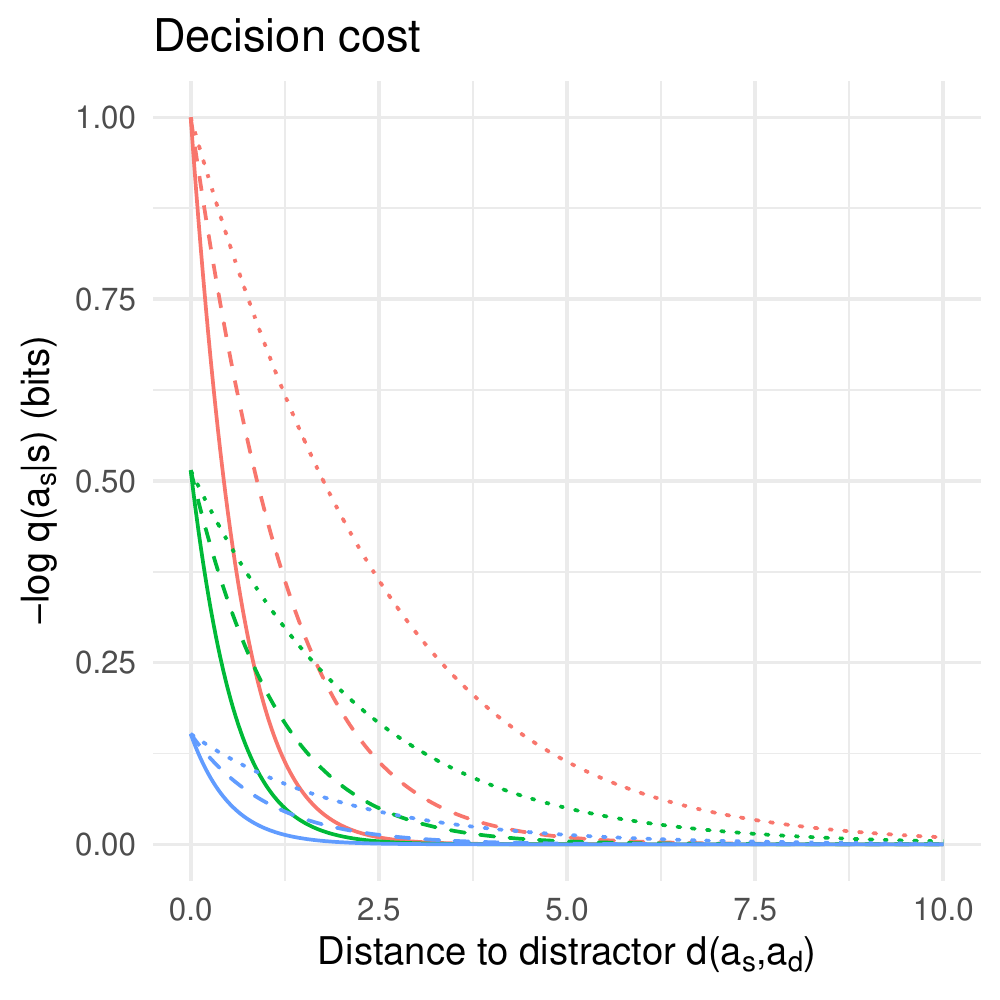}
        \includegraphics[width=.38\textwidth]{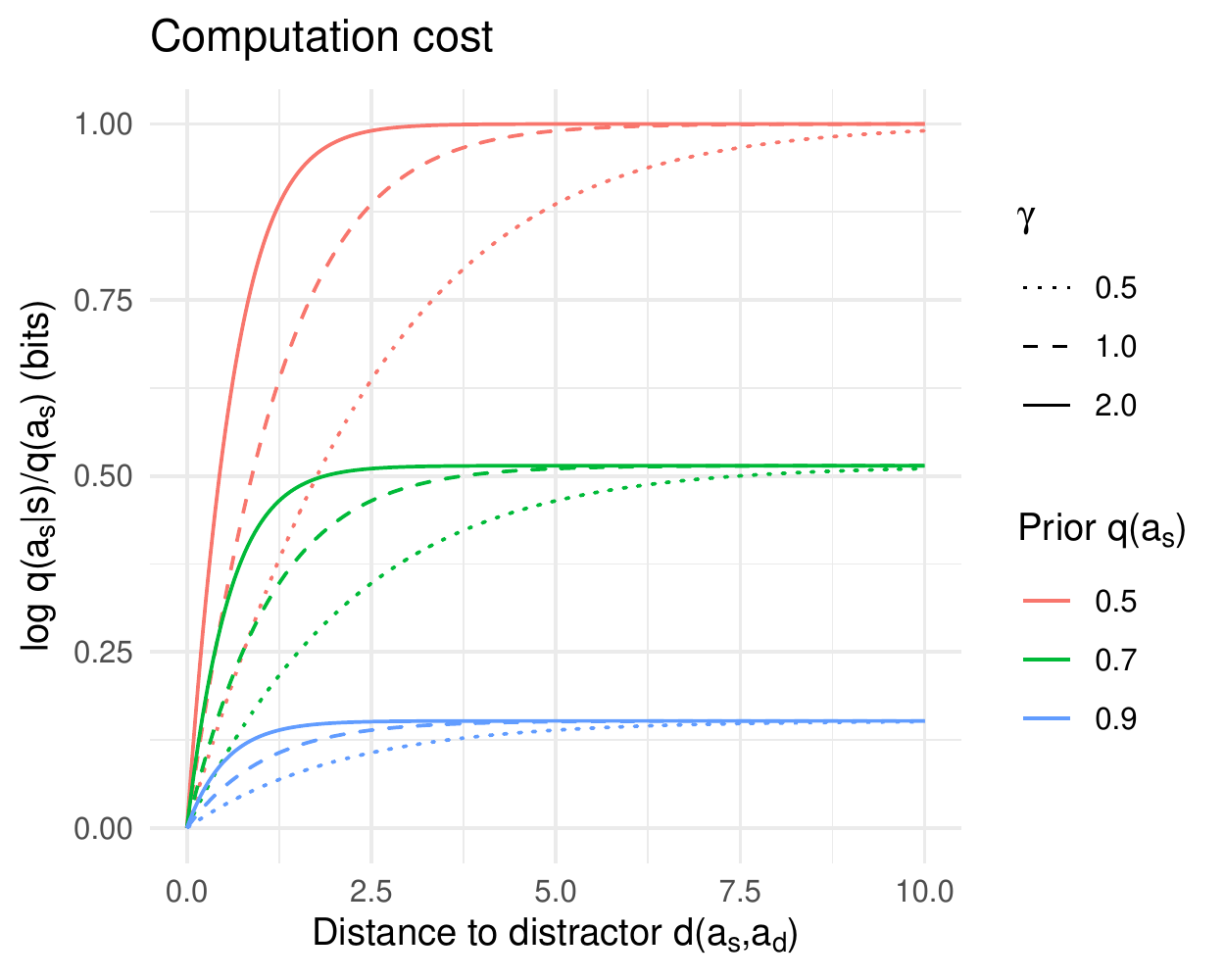}
    \caption{Interference between a target action $a_s$ and a distractor $a_d$ as a function of the semantic distance between them, $d(a_s, a_d)$, for varying values of resouce parameter $\gamma$ and the a priori probability $q(a_s)$. \textbf{Left}, the probability $q(a_s|s)$ of taking the appropriate action $a_s$ in state $s$. Note that when $d(a_s, a_d)=0$, we get $q(a_s|s)=q(a_s)$. \textbf{Center}, the decision cost $-\log q(a_s|s)$, which is high when $a_s$ and $a_d$ have low semantic distance. \textbf{Right}, the computation cost $\log \frac{q(a_s|s)}{q(a_s)}$.}
    \label{fig:logistic-interference}
\end{figure}

More formally, let the control objective be
\begin{equation}
    \label{eq:state-lagrangian}
    \mathcal{L}_q = \Big\langle d(a_s, a) \Big\rangle + \frac{1}{\gamma} I[S:A].
\end{equation}
This equation expresses that the agent will try to minimize the average distance between the selected action $a$ and the target action $a_s$, subject to a computation cost of $\frac{1}{\gamma}$ units per bit of information shared between $S$ and $A$.
Then following the logic in Eq.~\ref{eq:equilibrium}, the boundedly rational policy has the form
\begin{align}
    \label{eq:target-equilibrium}
    q(a|s) &= \frac{1}{Z(s)} q(a) \exp{- \gamma d(a_s, a)} \\
    \nonumber
    Z(s) &= \sum_a q(a) \exp{- \gamma d(a_s, a)}.
\end{align}
This policy will generally exhibit interference effects in $d$. To simplify the presentation, consider a scenario where there are only two possible actions given a state $s$: the target action $a_s$ and the distractor $a_d$. Plugging in to Eq.~\ref{eq:target-equilibrium}, we find that the probability of the target action $a_s$ in state $s$ is given by a logistic curve:\footnote{The probability of the target action $q(a_s|s)$ is calculated as follows:
\begin{align}
    \nonumber
    q(a_s|s) &= \frac{q(a_s) \exp{-\gamma d(a_s, a_s)}}{q(a) \exp{-\gamma d(a_s, a_d)} + q(a_d) \exp{-\gamma d(a_s, a_d)}} \\
    \nonumber
    &= \frac{q(a_s) \exp{0}}{q(a_s) \exp{0} + q(a_d) \exp{-\gamma d(a_s, a_d)}} \\
    \nonumber
    &= \frac{q(a_s)}{q(a_s) + q(a_d) \exp{-\gamma d(a_s, a_d)}} \\
    \nonumber
    &= \frac{1}{1 + \frac{q(a_d)}{q(a_s)} \exp{-\gamma d(a_s, a_d)}}.
\end{align}
This is an instance of the general logistic curve
\begin{equation*}
    f(x) = \frac{1}{1 + \exp{-k(x - x_0)}}
\end{equation*}
with slope parameter $k=\gamma$ and initial condition $x_0 = \frac{1}{\gamma}\log \frac{q(a_d)}{q(a_s)}.$
More generally, given a set of distractors $a_d \neq a_s$, the probability of the correct action $a_s$ is
\begin{equation*}
    q(a_s|s) = \frac{1}{1 + \sum_{a_d \neq a_s} \frac{q(a_d)}{q(a_s)} \exp{-\gamma d(a_d, a_s)}}.
\end{equation*}
}
\begin{equation}
    \label{eq:logistic-interference}
    q(a_s|s) = \Bigg [ 1 + \frac{q(a_d)}{q(a_s)} \exp{-\gamma d(a_s, a_d)} \Bigg ] ^{-1}.
\end{equation}
The curve is illustrated in Figure~\ref{fig:logistic-interference}. The important part of Eq.~\ref{eq:logistic-interference} is the second term, which represents the effect of interference between the target action $a_s$ and the distractor action $a_d$. As this interference term gets larger, the value of $q(a_s|s)$ gets smaller. This interference term is large when (1) the distractor action $a_d$ is a priori likely, and (2) the distractor action $a_d$ is close to the target action $a_s$. 

An agent with a control objective as in Eq.~\ref{eq:state-lagrangian} will therefore show similarity-based interference in terms of errors in the action taken. This interference also manifests in decision cost for action $a_s$:
\begin{align}
    \nonumber
    \text{Decision cost} &\sim -\log q(a_s|s) \\
    \label{eq:control-cost-simple}
    &= \log \Big ( 1 + \frac{q(a_d)}{q(a_s)} \exp{ -\gamma d(a_s, a_d) } \Big ).
\end{align}
This function decreases as $d(a_s, a_d)$ increases. The shape of this function is shown in Figure~\ref{fig:logistic-interference}.

Applying this logic to word production, we predict interference effects among semantically similar production targets when both are likely actions given the agent's state. Consider a state where a person sees a picture of an apple, and the words ``apple'' and ``pear'' are both a priori likely for some reason. This corresponds to target action $a_s = \texttt{say ``apple''}$ and distractor action $a_d = \texttt{say ``pear''}$, with $q(a_s)$ and $q(a_d)$ both high, and $d(a_s, a_d)$ low. A boundedly rational agent will erroneously say ``pear'' in this state more often than if the distractor were something less similar, such as $a_d^\prime = \texttt{say ``car''}$; furthermore, the action $a_s = \texttt{say ``apple''}$ can only be produced at higher decision cost due to the presence of the distractor. The reason is that when the distractor is `car', the relevant distance is $d(a_s, a_d^\prime) \gg d(a_s, a_d)$, leading to a lower probability of confusion in the action policy, and a lower decision cost. We can also see that the probability of an erroneous production and the decision cost will increase as the distractor's a priori probability $q(a_d)$ increases.

This example embodies the core logic of the RDC account of semantic interference. Below, I will demonstrate this logic in a more thoroughly worked out model of the Stroop / Picture--Word Interference Task. That simulation will require a more involved control model, but the underlying cause of similarity-based interference remains the same as in this example. 


\section{Model of Picture--Word Interference}
\label{sec:pwi-simulation}

Here, I show that RDC can capture some of the major characteristics of semantic interference in the Stroop task and the Picture--Word Interference task. 

\subsection{The phenomena}

\begin{figure}
    \centering
    \includegraphics[trim={2cm 6cm 2cm 12cm},clip,scale=.2]{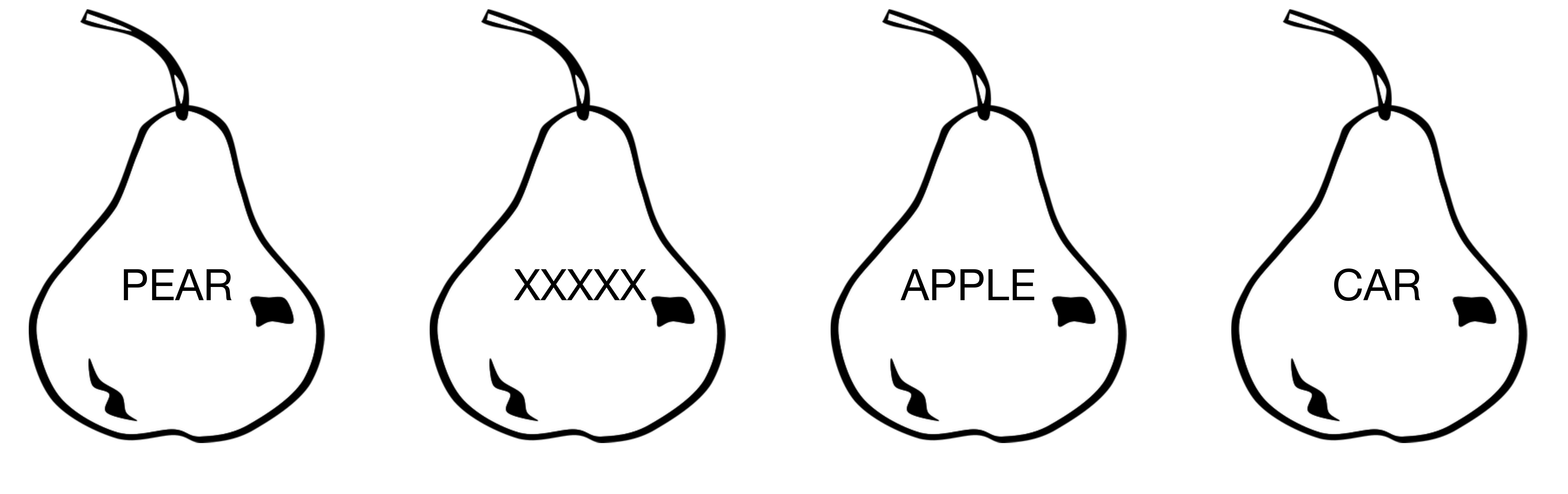}
    \caption{Conditions of a Picture--Word Interference experiment. From left to right: the \key{congruent}, \key{neutral}, \key{incongruent}, and \key{unrelated} conditions (see text).}
    \label{fig:pwi-example}
\end{figure}

\key{Picture--Word Interference} (PWI) is one of the most well-studied phenomena in language production and cognitive control \citep{burki2020what}. The task evokes similarity-based interference in picture naming by superimposing a text word over an image, and asking a subject to name the image \citep{lupker1979semantic}. Examples are shown in Figure~\ref{fig:pwi-example}. The \key{Stroop task} is closely related \citep{stroop1935studies,macleod1991half,vanmaanen2009stroop,starreveld2017picture}: in this task, a word such as \textbf{\textcolor{red}{BLUE}} is presented in red ink, and subjects are asked to name the color of the ink.

The hallmark PWI effect is that subjects are slower to name the image in the presence of a superimposed word which is semantically categorically related to the image (the \emph{incongruent} condition in Figure~\ref{fig:pwi-example}), as compared to their reaction times when the superimposed text is a neutral string such as \texttt{XXXXX} (the \emph{neutral} condition in Figure~\ref{fig:pwi-example}). Furthermore, reaction times are fastest when the superimposed word is the same as the name of the image (the \emph{congruent} condition), and if the superimposed text is a semantically unrelated word (the \emph{unrelated} condition), reaction times are somewhere between the neutral and incongruent conditions. 

Many PWI and Stroop experiments include only a neutral or an unrelated baseline condition, rather than all four of these conditions, which has resulted in some variance in terms of the size of the reported interference effect \citep{macleod1991half}. The neutral and unrelated conditions are referred to together as the \emph{baseline} condition.

\subsection{Related work}

Because of its empirical robustness and (apparent) conceptual simplicity, PWI and Stroop tasks have been the target of many computational cognitive models throughout the past three decades, and subject to intense controversies about the mechanism that gives rise to the observed interference effect.

The main controversy in the literature is over whether PWI effects are driven by a competitive process during lexical selection, where multiple responses are competing for priority, resulting in slowdown \citep{roelofs1992spreading,levelt1999theory,damian2003locus,belke2005age,abdelrahman2009semantic} or by the need to exclude the distractor from an articulatory buffer \citep[for example,][]{mahon2007lexical}. 

The most extensively documented model of PWI is WEAVER++ \citep{levelt1999theory,roelofs2003goal}, a model of word production based on production rules where similarity-based interference emerges due to competition in lexical selection. Alternative models include the Swinging Lexical Network \citep{abdelrahman2009semantic}.

In contrast to these existing models, the RDC account of interference in word production is a computational-level model which works by specifying only the problem that is being solved by the cognitive system, without making any commitments to algorithmic-level details \citep{marr1982vision}. The theory and its assumptions are specified completely by (1) the control objective, which is the problem that the cognitive system is trying to solve, and (2) the linking function from cognitive costs to observables such as RT. 

As we will see, the control objective that reproduces PWI effects specifies only that there is some computational bottleneck involved in integrating information from bottom-up sensory input and top-down behavioral goals---whether this bottleneck happens in lexical selection, articulation, etc. is unspecified. The computational bottleneck might arise mechanistically due to dynamics of spreading activation, competing production rules, etc. The question of whether the interference effect arises because of competition or response exclusion does not arise at this level of abstraction.

I am aware of two previous information-theoretic models of the Stroop task. \citet{zenon2019information} presented a model of information-processing costs in the Stroop task which predicts that performing an unusual goal (i.e., naming a picture rather than reading a word) results in increased difficulty. Their model does not use bounded-optimal policies. Also, \citet{christie2019information} models the RT response distribution for congruent, incongruent, and neutral trials in a Stroop task using an information-theoretic model in which conflicting control signals are superposed and must be decoded at high cost. Their model involves a policy which receives noisy bottom-up and top-down signals and must decide on an action in the presence of uncertainty due to this noise. While their model is based on noise, rather than lossy compression, it is fundamentally similar to the model presented here because it involves rational action under cognitive constraints. The RDC model presented here explains more of the phenomena associated with PWI than these previous models, including the higher interference for semantically related distractors, the relative size of facilitation and interference effects, and the emergence of a small interference effect when reading words after many congruent trials.

\subsection{RDC account}

A full model of PWI requires a more complex setup than the simple interference example above. In particular, whereas the interference model given by Eq.~\ref{eq:state-lagrangian} involved a policy conditional only on an input state, a full model of PWI requires a policy conditional on \emph{two} inputs: a perceptual state and a top-down behavioral goal. 

To model PWI, let $G$ be a random variable representing a speaker's top-down goals, i.e. whether the goal is to name a picture/color or to read a word. That is, $G$ is a random variable taking values in $\{\texttt{name}, \texttt{read}\}$. Let $S$ be a random variable representing a speaker's perceptual state---that is, the particular word and picture that the speaker is looking at. A speaker implements a production policy $q(a|g,s)$ subject to information-processing costs.

In order to capture semantic interference in PWI, it turns out that the minimal necessary informational constraint in the control objective is on the \key{synergy} of the three variables $G$, $S$, and $A$---the extent to which producing action $A$ requires information that is \emph{only} present in the combination of $G$ and $S$, but not present in either alone \citep{mcgill1955multivariate,bell2003coinformation}. Synergy is defined as:
\begin{equation}
    \label{eq:synergy}
    S[G:S:A] = I[G:A|S] - I[G:A],
\end{equation}
with the \key{conditional mutual information} $I[G:A|S]$ defined as
\begin{equation*}
    I[G:A|S] = \Bigg \langle \log \frac{q(a|g,s)}{q(a|s)} \Bigg \rangle.
\end{equation*}
Therefore, the synergy can also be written as
\begin{equation}
    \label{eq:synergy-average}
    S[G:S:A] = \Bigg \langle \log \frac{q(a|g,s)q(a)}{q(a|g)q(a|s)} \Bigg \rangle,
\end{equation}
suggesting that the computation cost for a particular action $a$, goal $g$, and perceptual state $s$ is simply the value inside the average in Eq.~\ref{eq:synergy-average}.

The value of synergy can be negative, indicating a scenario where random variables $G$ and $S$ both \emph{redundantly} contain some overlapping information about $A$. However, in the case of PWI, because of the way the problem is set up, Eq.~\ref{eq:synergy} is nearly always positive. This is because the quantity $I[G:A]$ is likely to be negligible in the case of PWI---because knowing only that your goal is to read or name something provides almost no information about the appropriate action unless you also know \emph{which} word you are supposed to read or name. 

Let $d : A \times A \rightarrow \mathbb{R}^{(+)}$ be a semantic distance measure on production actions $A$, as defined in Eq.~\ref{eq:cost-diff}.
Then the speaker's production policy is a minimum of the policy objective:
\begin{equation}
    \label{eq:synergy-penalty}
    \mathcal{L} = \Big \langle d(a^g_s, a) \Big \rangle + \frac{1}{\gamma} S[G:S:A],
\end{equation}
where $a^g_s$ indicates the correct action to be taken in state $s$ with goal $g$.
The minima of the policy objective in Eq.~\ref{eq:synergy-penalty} have the form:
\begin{equation}
    \label{eq:synergy-policy}
    q(a|g,s) = \frac{q(a|s)q(a|g)}{Z(g,s) q(a)} \exp{-\gamma d(a^g_s, a)}.
\end{equation}

Below, I will first analyze the policy in Eq.~\ref{eq:synergy-policy} and show that it demonstrates semantic interference under reasonable default parameter settings in a simulation of the PWI task, and that it can capture some of the major qualitative empirical patterns observed in PWI studies when we vary the parameters of the simulation. 

\subsection{Simulation}

I model the basic Stroop task with the following setup. An agent has access to a behavioral goal and a perceptual state, and produces an output action in response to these. The perceptual state consists of a picture and a written word. The behavioral goal specifies whether the agent should read the word or name the picture. Each word and each picture is associated with a single appropriate target action. I assume a semantic distance metric among output actions such that there is a constant cost for taking an action $a$ when the target action was $b$; this cost function does not depend on the behavioral goal. 

In more detail: the behavioral goal is a random variable $G$ that can take one of two values, $g \in \{\texttt{name}, \texttt{read}\}$, with the probability of the goal being \texttt{name} equal to a parameter $p_\texttt{name} = \frac{1}{10}$. This low probability is meant to reflect the low frequency of the goal of naming a picture as opposed to reading a word in everyday experience. The perceptual state is represented by the random variable $S$ and takes values in \emph{pairs} of discrete objects $\langle w, p \rangle$, representing a state where an agent is seeing word $w$ superimposed on picture $p$. The number of possible words is $N_w$ and the number of possible pictures is $N_p$; in the simulations below, unless otherwise noted, I fix $N_w = N_p = 10$ and assume a uniform distribution on the possible states. The output actions are represented by a random variable $A$ taking one of $N_a = 10$ different values. Each goal $g$ and state $s$ is associated with a target action $a^g_s$ defined as follows: given the goal $g=\texttt{read}$ and the state $s=\langle w, p \rangle$, the target action is $w$; given the goal $g=\texttt{name}$, the target action is $p$. The distance metric among output actions $d : A \times A \rightarrow \mathbb{R^{(+)}}$ varies linearly from 0 (for identical actions) to 10 (for maximally distinct actions).

The last parameter we need to specify an RDC policy is the scalar $\gamma$, which gives the computational resources (inverse cost) available for synergistic information processing in the model. With all these parameters in hand, we can compute the RDC policy from the control objective in Eq.~\ref{eq:synergy-penalty}. Simulation parameters are summarized in Table~\ref{tab:simulation-parameters}. 

\begin{table}[]
    \centering
    \begin{tabular}{@{}lll@{}}
        \toprule
        Parameter & Value & Meaning \\
        \midrule
         $p_{\texttt{name}}$ & $\frac{1}{10}$ & The a priori probability of the behavioral goal being to name, rather than read. \\
         $N_w$ & 10 & Number of different words in possible perceptual states. \\
         $N_p$ & 10 & Number of different pictures in possible perceptual states. \\
         $\gamma$ & 3 & Resources for synergistic information processing (see Eq.~\ref{eq:synergy-penalty}). \\
         \bottomrule
    \end{tabular}
    \caption{Parameters of the simulation of the Stroop task. See text for detailed discussion.}
    \label{tab:simulation-parameters}
\end{table}

As a more concrete example, suppose the goal $g=\texttt{name}$, and the perceptual state is the pair $\langle \texttt{apple}, \texttt{pear} \rangle$, representing the word ``apple'' superimposed on a picture of a pear. Because the goal is $g=\texttt{name}$, the target action $a^g_s$ is to say ``pear.'' If the agent takes this action, then the distortion is zero, because $d(\texttt{pear}, \texttt{pear}) = 0$. On the other hand, if the agent takes the action of saying ``apple'', then the distortion is $d(\texttt{pear}, \texttt{apple})$, which may be small, since these are semantically related words that share many features. Because this distortion is low, an agent may be attracted toward saying ``apple'', which has higher distortion than ``pear'', but has lower computation cost because it does not require attending to the costly behavioral goal. Then the probability of producing the correct word ``pear'' will be low and the decision cost for the correct word ``pear'' will be high.

Given a state $\langle w, p \rangle$ and a goal $g$, we can define one part of the state as the `target' and another as the `distractor'. When $g=\texttt{name}$, the target is $p$ and the distractor is $w$. When $g=\texttt{read}$, the target is $w$ and the distractor is $p$. In each state, there will be a certain semantic distance between the target and distractor, called the \key{distractor distance}. If $a(w)$ represents the action associated with $w$ and $a(p)$ is the action associated with $p$, then when $g=\texttt{name}$, the distractor distance is $d(a(p), a(w))$; when $g=\texttt{read}$, the distractor distance is $d(a(w), a(p))$. 

The major conditions of PWI experiments are the congruent, incongruent, neutral, and unrelated conditions (defined in Figure~\ref{fig:pwi-example}). So far, we have the ability to model three of these: the congruent condition corresponds to the case where the distractor distance is 0 (i.e., the target actions are identical across goals: $a(w) = a(p)$); the incongruent condition corresponds to the case where distractor distance is low; and the unrelated condition means the distractor distance is high. I will return to the neutral condition below. 

\section{Results}

\subsection{Basic results}

\begin{figure}
    \centering
    \includegraphics[width=\textwidth]{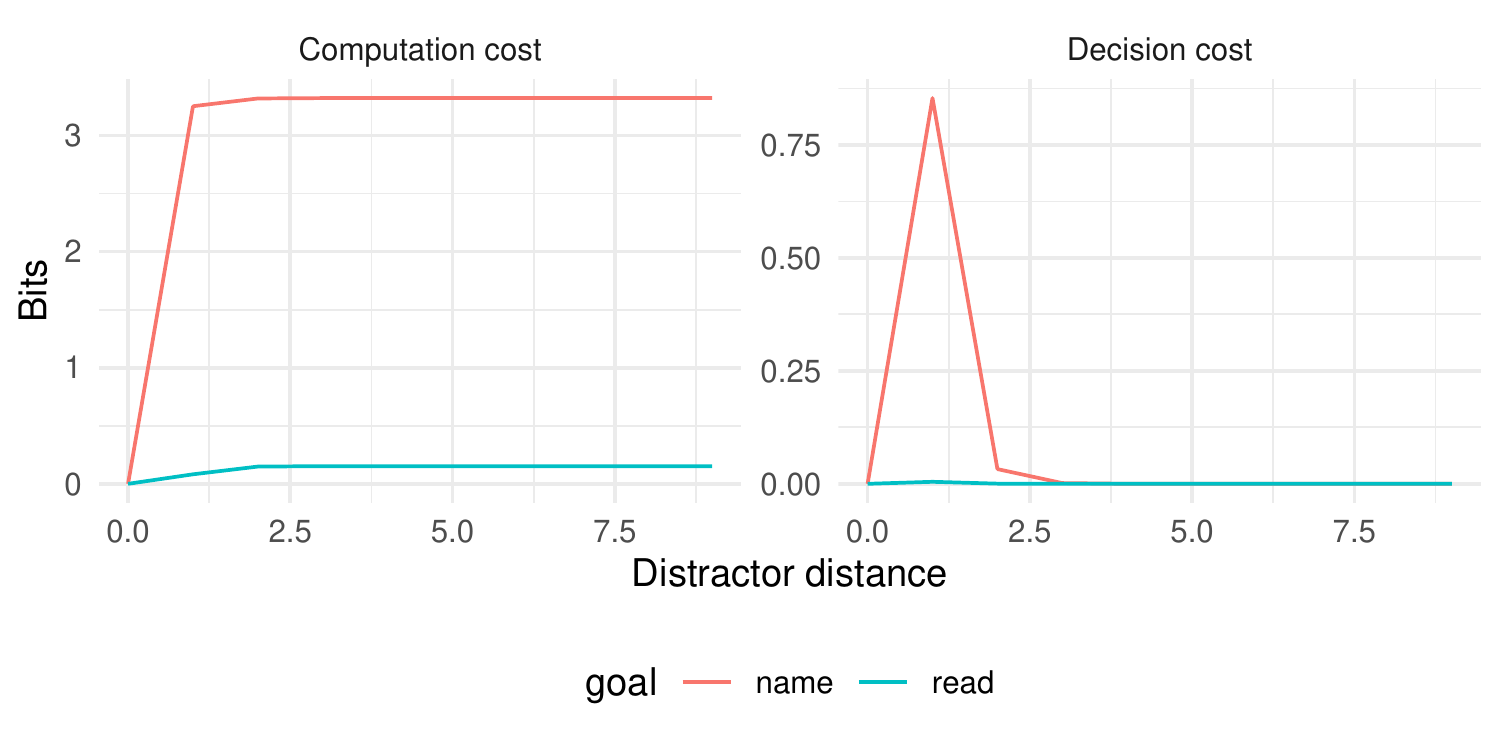}
    \caption{Simulated costs in Picture--Word Interference task, as a function of semantic distance between target and distractor.}
    \label{fig:simple-simulation}
\end{figure}

In Figure~\ref{fig:simple-simulation}, I show the decision cost and the computation cost based on the simulation with the default parameter values, as a function of distractor distance. We see a few basic patterns:
\begin{itemize}
    \item There is no decision cost and low computation cost when $d=0$, corresponding to the congruent condition in experiments. 
    \item Semantic interference exists in the decision cost. The interference is high for close words (corresponding to the incongruent condition), and falls off rapidly at distant words (corresponding to the unrelated condition). 
    \item When the goal is $g=\texttt{read}$, interference of any kind is negligible.
\end{itemize}

Computation cost is essentially a constant function of the goal, except when the appropriate actions given the two goals coincide (distractor distance $0$). In fact, for goal $g$ and state $s=\langle w,p \rangle$, we can derive a simple approximate expression for the computation cost:\footnote{The approximation follows from setting $q(a|g,s) = \delta_{aa^g_s}$ and calculating computation cost as $\log \frac{q(a|g,s)q(a)}{q(a|g)q(a|s)}$.}
\begin{equation}
    \label{eq:comp-cost-approx}
    \text{Computation cost} \sim \begin{cases}
    0 \text{ if } a(w) = a(p) \\
    \log \frac{1}{p(g|s)} \text{ otherwise}
    \end{cases}.
\end{equation}
Eq.~\ref{eq:comp-cost-approx} indicates a constant cost associated with rare behavioral goals, proportional to the log probability of that goal. This approximation recovers the model of Stroop interference from \citet{zenon2019information}.

This most basic simulation already captures several qualitative patterns from the empirical literature \citep[as listed by][]{macleod1991half}. First, we recover the fact that naming is generally slower than reading \citep{cattell1886time}, as indicated by the uniformly higher computation cost for naming. Second, we recover the existence of facilitation in the congruent condition, reflected in lower decision cost and lower computation cost when distractor distance is zero. Third, we recover the existence of interference in the incongruent condition relative to the congruent condition and the unrelated condition, as reflected in the decision cost. Fourth, interference exists for the naming task but is negligible in the reading task. Fifth, the interference effect is gradient \citep{klein1964semantic}: when the distractor is \emph{more} semantically similar to the target, there is more interference; this is reflected in the decision cost for the naming condition, which is high at distance 1, lower at distance 2, and still lower at distance 3.

The semantic gradient deserves a bit more discussion. There has been controversy in the literature on Picture--Word Interference about whether a semantic gradient really exists, as opposed to a categorical effect for distractors that are in the same category as the target \citep{hutson2014semantic,burki2020what}.  In the RDC model, there is a semantic gradient observable in the decision cost, but it falls off very rapidly from distance 1 to distance 2, and distance 2 shows only barely more interference than distance 3. Therefore the theory predicts that a semantic gradient does exist, but it is highly concentrated, and thus might be hard to detect in experiments. 


Above, I have shown that RDC can capture the basics of semantic interference in PWI tasks in a simulation with simple and reasonable default parameter settings. Next, I will show how we can recover more of the empirical patterns by varying the parameters of the simulation and the model.


\subsection{Neutral versus unrelated trials}
\label{sec:neutral}

The PWI task has a fourth major condition: the \emph{neutral} condition, where a picture is presented along some kind of neutral orthographic stimulus that would not reasonably be read out loud, such as \texttt{XXXXX}. Here, I will incorporate this condition into the simulation and show that we immediately recover three empirically-attested patterns: (1) there is facilitation in the congruent condition relative to the neutral condition, (2) there is interference in the unrelated condition relative to the neutral condition, and (3) the size of facilitation is small relative to the size of interference \citep{macleod1991half}.

Recall that in the basic simulation, the a priori probability that the behavioral goal is $g=\texttt{name}$ rather than $g=\texttt{read}$ is $\frac{1}{10}$. I model the neutral condition by adding into the simulation a set of states $s_\text{neutral}$ where $p(g=\texttt{name} | s_\text{neutral}) = \frac{9}{10}$, modeling the scenario where a subject sees \texttt{XXXXX} superimposed on an image. The idea is that given such a state, a subject would only expect to actually read the stimulus (saying ``eks eks eks eks eks'') $\frac{1}{10}$ of the time. Outside of a state with a neutral distractor $s_\text{neutral}$, the probability of naming is still $\frac{1}{10}$.

\begin{figure}
    \centering
    \includegraphics[width=\textwidth]{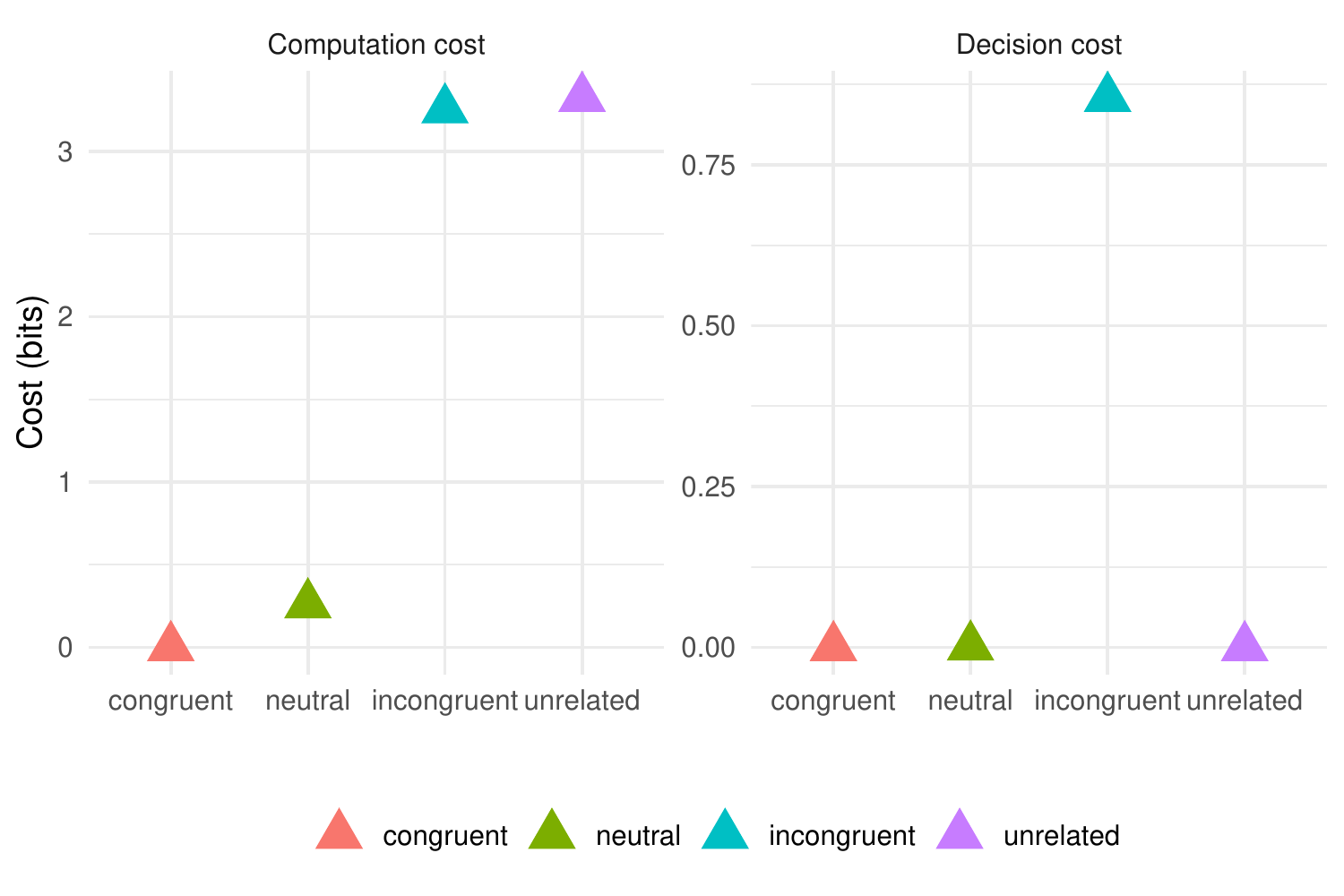}
    \caption{Simulated costs by PWI task condition. See text for definitions.}
    \label{fig:neutral-simulation}
\end{figure}

Figure~\ref{fig:neutral-simulation} shows the simulated decision and computation costs for four experimental conditions: congruent (the case where the distance $d=0$), incongruent (the case where $d=1$), unrelated (the case where $d=9$), and neutral (the case where $s=s_\text{neutral}$). 

The three empirical patterns are captured here by the computation cost. The neutral condition has drastically reduced computation cost relative to the semantic and unrelated conditions, indicating facilitation. Also, the computation cost is slightly less in the congruent case relative to the neutral case, indicating facilitation. Also, the size of the facilitation effect (the difference between neutral and congruent conditions) is small relative to the interference effect (the difference between neutral and incongruent conditions). 

The model robustly recovers the existence of facilitation and interference, and the relative magnitude of facilitation and interference depends on a model parameter: the probability $p(g=\texttt{name} | s=s_\text{neutral})$.\footnote{The default values for $p(g|s)$ have not been tuned to fit the human data, but were selected a priori and kept constant throughout all simulations.} Therefore, it is therefore possible to make a prediction: the facilitation effect should get larger under any manipulation that makes the orthographic string in the neutral condition more and more like something that someone would reasonably read. In fact, there is already some evidence in this direction in the literature: pseudowords, which presumably fall somewhere between \texttt{XXXXX} and a real word in terms of $p(g=\texttt{name} | s)$, cause less interference than real words in the Stroop task \citep{klein1964semantic}.

\subsection{Fit to human data}

Here I relate the simulated computation and decision costs to empirical human RT data. To do so, we need a more specific linking function from computation and decision cost to RT. 

I propose that RT can be predicted from a linear combination of computation and decision cost. That is, the predicted RT in a condition is given from cognitive costs by an affine transformation:
\begin{equation}
    \text{RT} = \alpha + \beta X + \gamma Y,
\end{equation}
where $X$ is computation cost, $Y$ is decision cost, and $\alpha$, $\beta$, and $\gamma$ are non-negative scalars. This linking function supposes that computation cost and decision cost are each associated with some fixed rate of information processing, given by $\beta$ and $\gamma$ respectively, in terms of milliseconds per bit. The scalar $\alpha$ represents a constant RT delay across conditions.

\begin{figure}
    \centering
    \includegraphics[width=\textwidth]{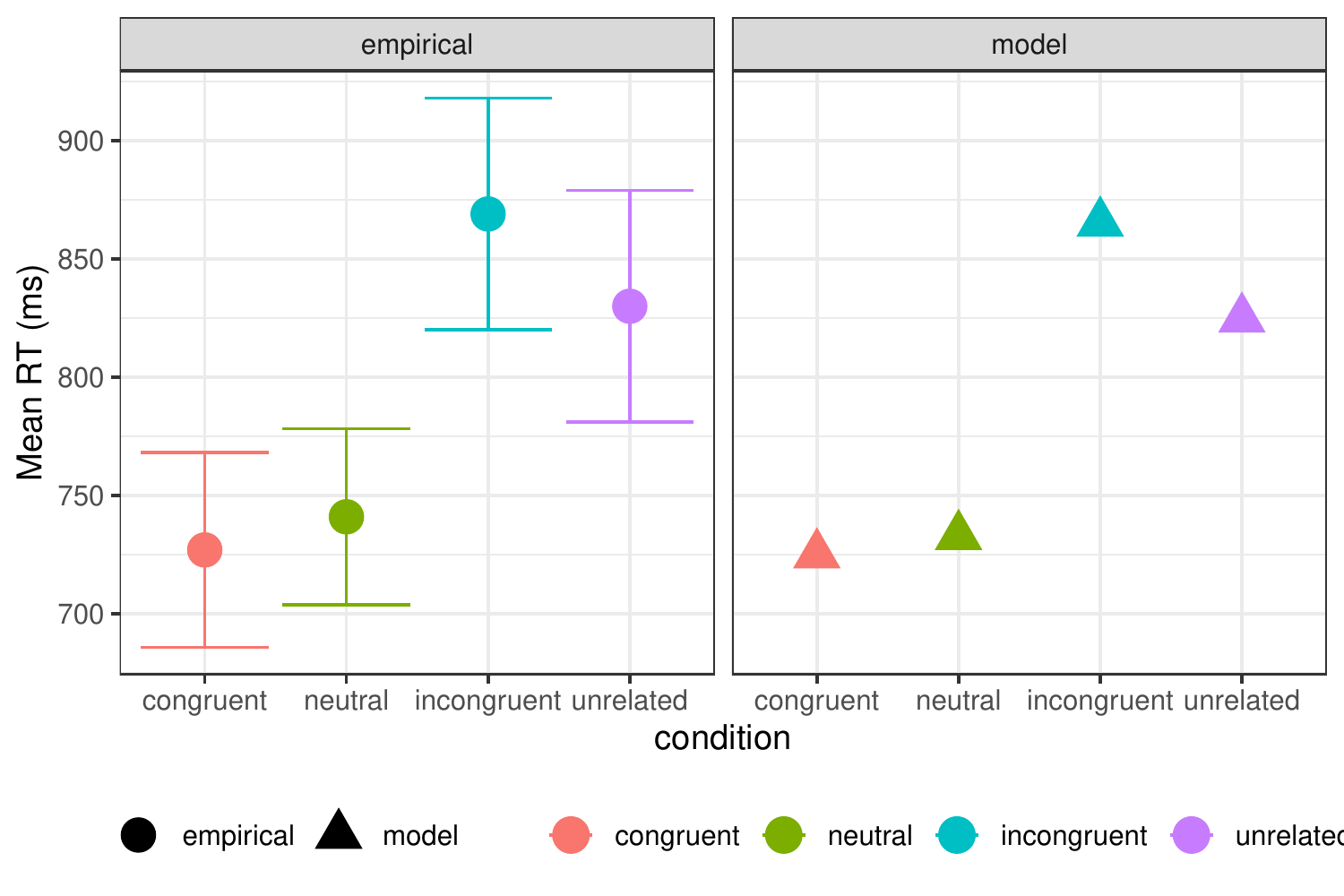}
    \caption{Empirical mean RTs for PWI conditions from \citet{roelofs2017distributional}, compared with model predictions ($\text{RT} = 730 + 30 \times \text{Computation cost} + 50 \times \text{Decision cost}$). Error bars show 95\% confidence intervals of the mean in the empirical data.}
    \label{fig:rp-mean}
\end{figure}

Figure~\ref{fig:rp-mean} shows a comparison of empirical mean RTs in a PWI task, drawn from \citet{roelofs2017distributional}, compared against simulated RTs, with $\alpha = 725$ ms, $\beta = 30$ ms/bit, and $\gamma=50$ ms/bit.\footnote{All of the scaling factors presented in this section were derived by linear regression on the empirical means, followed by rounding. From the linear regressions, the optimal models before rounding are
\begin{align}
    \nonumber
    \text{mean RT} &\approx 730 + 30 \times \text{Computation cost} + 48 \times \text{Decision cost} \\
    \nonumber
    \mu &\approx 612 + 25 \times \text{Computation cost} + 23 \times \text{Decision cost} \\
    \nonumber
    \tau &\approx 119 + 3 \times \text{Computation cost} + 29 \times \text{Decision cost}.
\end{align}
} 
This mixture gives a good qualitative fit to the human data. 

The relationship of information-processing costs to RT may not be so simple as an affine transformation. In particular, RT distributions appear to follow what is called an ex-Gaussian distribution \citep{ratcliff1979group,luce1986response,balota2008beyond}. Briefly, an ex-Gaussian distribution is the sum of a Gaussian distribution with mean $\mu$ and an exponential distribution with rate $\tau$. The resulting distribution is skewed toward the right when compared with a Gaussian distribution. Interestingly, the $\mu$ and $\tau$ parameters of the ex-Gaussian distribution might reflect different aspects of cognitive processing. The $\mu$ and $\tau$ components of RT distributions from PWI and Stroop tasks have been studied in previous work \citep{heathcote1991analysis,mewhort1992response,spieler2000levels,roelofs2012attention,piai2011semantic,piai2012distractor,scaltritti2015distributional}. 

\begin{figure}
    \centering
    \includegraphics[width=\textwidth]{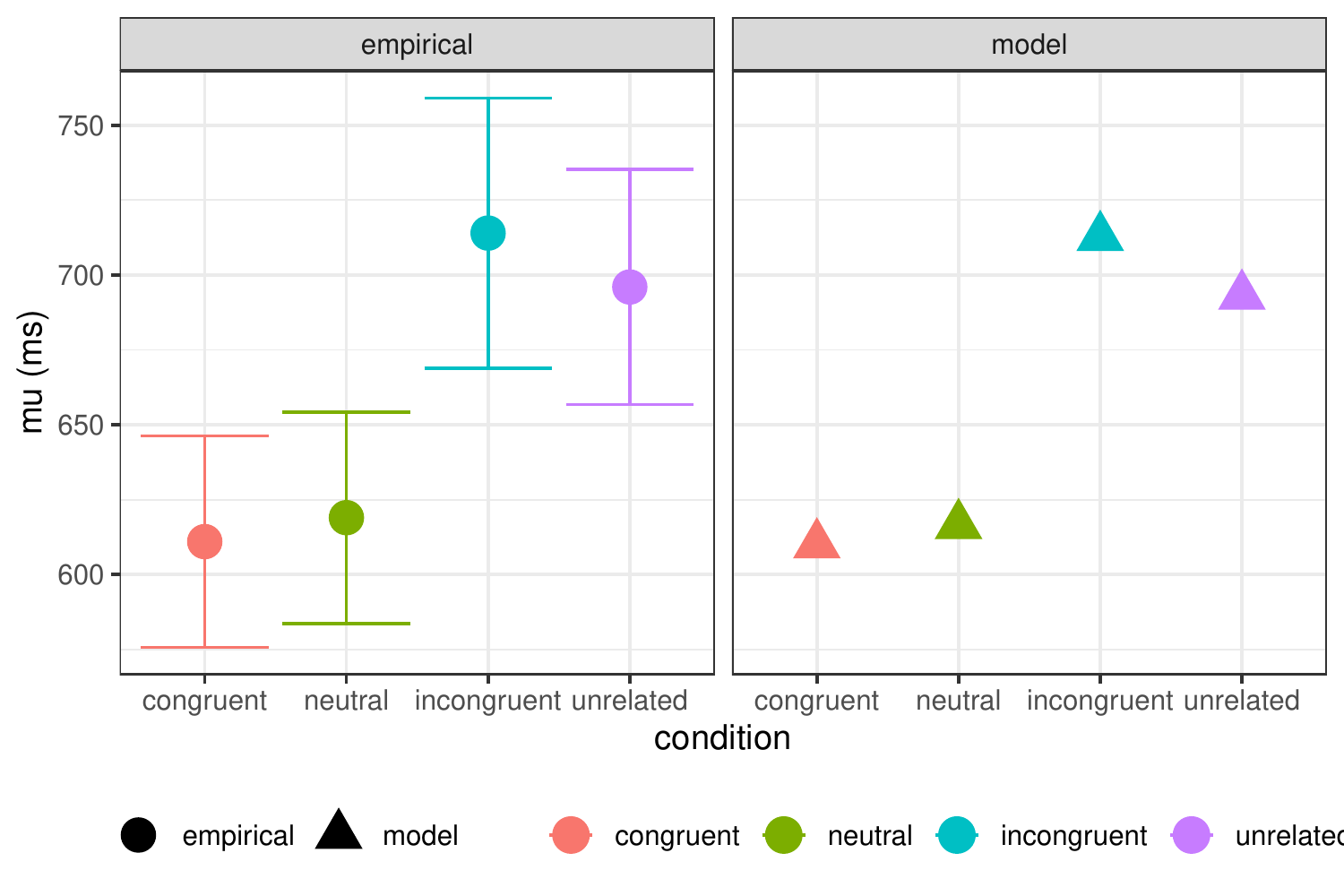}
    \caption{Empirically estimated $\mu$ parameter of ex-Gaussian RT distribution for PWI conditions from \citet{roelofs2017distributional}, compared with model predictions ($\mu = 610 + 25 \times \text{Computation cost} + 25 \times \text{Decision cost}$).}
    \label{fig:rp-mu}
\end{figure}

\begin{figure}
    \centering
    \includegraphics[width=\textwidth]{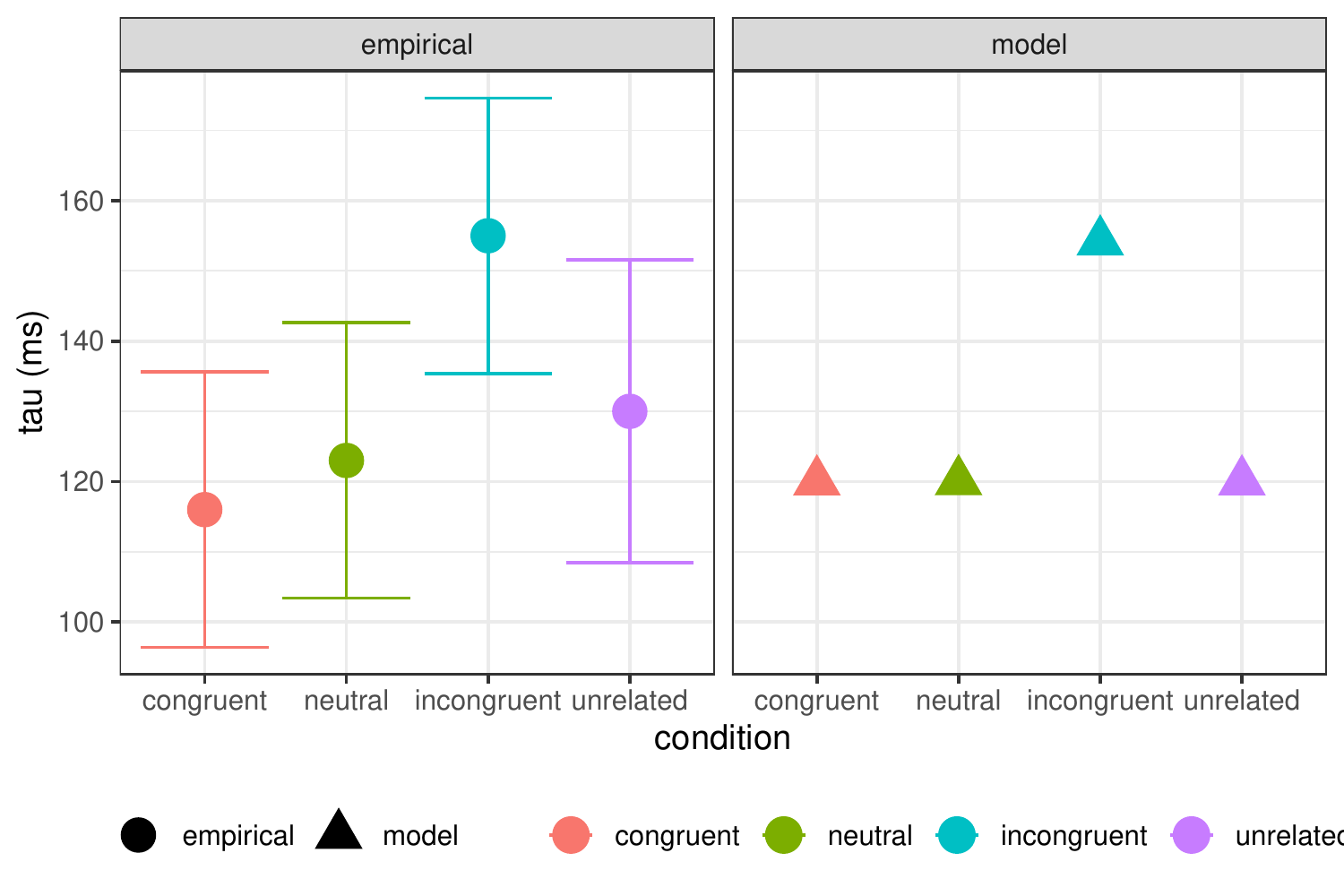}
    \caption{Empirically estimated $\tau$ parameter of ex-Gaussian RT distribution for PWI conditions from \citet{roelofs2017distributional}, compared with model predictions ($\tau = 120 + 40 \times \text{Decision cost}$).}
    \label{fig:rp-tau}
\end{figure}

Here I present an analysis comparing computation and decision costs to the full ex-Gaussian analysis of experimental PWI data, including congruent, incongruent, neutral, and unrelated conditions, performed by \citet{roelofs2017distributional}.
In Figure~\ref{fig:rp-mu}, I show their estimates of the $\mu$ parameter compared with an equal combination of computation cost and decision cost ($\alpha = 610$ ms, $\beta = 25$ ms/bit, $\gamma = 25$ ms/bit).\footnote{These values are intriguingly close to \citet{fan2014information}'s claim that information processing in the prefrontal cortex happens at $\sim 20$ ms/bit.} In Figure~\ref{fig:rp-tau}, I compare their $\tau$ estimates to decision cost alone ($\alpha = 120$ ms, $\beta=0$, $\gamma=40$ ms/bit). 
The reasonable qualitative match suggests that both computation and decision cost are reflected in the $\mu$ component of the RT distribution, while only decision cost is reflected in the $\tau$ component.
This would be in line with the pattern reported by \citet{roelofs2017distributional}: $\mu$ shows a contrast among neutral, unrelated, and incongruent conditions, while $\tau$ shows a contrast only between the incongruent condition and the others \citep[see also][who claim that the additional interference for semantically related distractors manifests only in $\tau$]{scaltritti2015distributional}.




\subsection{Reverse Stroop}

The \key{reverse Stroop effect} refers to a reversal in the difference between naming and reading in a Stroop task. Usually, interference happens in the naming task and not in the reading task. However, after a great deal of practice with color naming, subjects begin to show an interference effect in reading as well as naming, and the interference effect in naming shrinks \citep{stroop1935studies,macleod1991half}. 

The reverse Stroop effect is usually taken to result from greater practice and familiarity with the naming task. In other words, with practice and exposure, the prior probability of the naming task $p_\texttt{name}$ increases. Therefore, I propose to model reverse Stroop by increasing the value of the parameter $p_\texttt{name}$ in the PWI simulation.

\begin{figure}
    \centering
    \includegraphics[scale=.8]{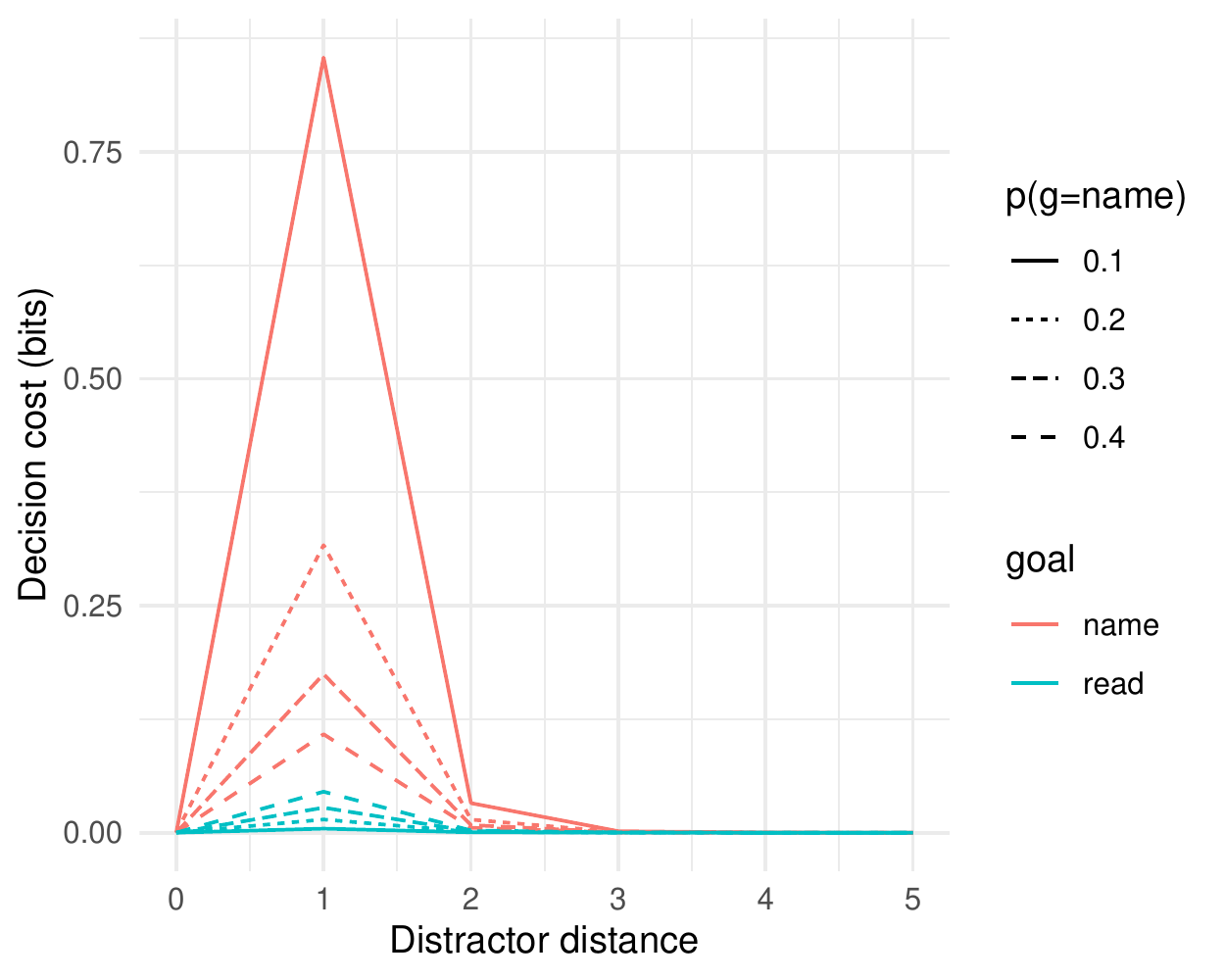}
    \caption{Decision cost for PWI under varying values of $p_\texttt{name}$. A reverse Stroop effect emerges in the cost under the reading goal.}
    \label{fig:varying-g}
\end{figure}

Figure~\ref{fig:varying-g} shows computation and decision costs under varying $p_\texttt{name}$. As this value increases, a reverse Stroop effect emerges, in that the reading task begins to show semantic interference in the decision cost. Meanwhile, the semantic interference associated with naming decreases (reflected in decision cost), and the general difficulty of the naming task decreases (reflected in computation cost).

\subsection{Further phenomena}

Here I briefly discuss how the RDC account of PWI can capture some additional phenomena from the literature. 

A great deal of work has claimed that interference is modified when the distractor word is not in the `response set' of the target word, i.e. when the distractor word is not a word that could be said in response to one of the pictures in a PWI experiment. There have been various claims of attenuation, nullification, or even reversal of semantic interference for words not in the response set in PWI \citep{roelofs1992spreading,piai2012distractor} \citep[but see][for claims that membership in the response set does not matter for semantic interference]{caramazza2000semantic,caramazza2001set}. The empirical picture remains complex, and appears to be highly dependent on stimulus-onset asynchrony. For a review and meta-analysis, see \citet{burki2020what}.

The RDC account as presented here predicts that interference is attenuated for distractors outside the response set. Recall the solution to the control objective from Eq.~\ref{eq:synergy-policy}:
\begin{equation}
    \tag{\ref{eq:synergy-policy}}
     q(a|g,s) = \frac{q(a|s)q(a|g)}{Z(g,s) q(a)} \exp{-\gamma d(a^g_s, a)}.
\end{equation}
Suppose the perceptual state $s=\langle w, p \rangle$ where $w$ is not in the response set for the experiment, or where $w$ is not a word that could plausibly be said in response to a picture. This means that $q(a(w)|g=\texttt{name})$ is relatively low, so the distractor $a(w)$ would get relatively little probability mass in the policy. Therefore we would see attenuated interference effects in the decision cost for distractors outside the response set.

Another effect from the literature comes from experiments which attempt to create word--word interference: when two printed words are predicted and one must be selected. The outcome is that interference exists, but without a semantic gradient \citep{roelofs2003goal}. This can be accommodated in the RDC framework via a goal-dependent distortion measure among actions in the following way.

The cost of misnaming a picture of an apple as a pear might reasonably show a semantic gradient, but the cost of mis-reading a word might not, for the following reason. While naming a picture of an apple as a pear could be considered an understandable mistake, naming the written word `apple' as anything other than `apple' is categorically incorrect. So if the semantic distortion measure among actions is consequentialist, as RDC claims, then we might expect a flat distortion metric when the goal is to read a word, and a gradient metric when the goal is to name a picture. In that case, we would predict interference but no semantic gradient for word--word naming.

\subsection{Discussion}

It is striking that the framework laid out here can successfully model many aspects of PWI, despite being developed nearly entirely for purposes other than cognitive modeling. Rate--distortion theory was developed purely as an abstract theory of lossy communication, and its application to control problems has primarily been confined to the computer science and robotics literature. 

Furthermore, RDC can model the general empirical patterns of the picture--word interference task with few free parameters. The  degrees of freedom in the specification of the model are (1) the distribution over goals and states, (2) the information-processing resource parameters used to define the control objective (the scalar $\gamma$), and (3) the similarity metric among actions. All of these degrees of freedom correspond to quantities that can be independently estimated, at least in principle. The distribution over goals and states is set by the frequency of goals and states in a person's everyday experience; the information-processing cost parameters are set by studies of cognitive difficulty; and the similarity metric among actions is determined by the relative cost of the consequences of confusing one action for another. Because of the small number of free parameters in the theory, it should be highly amenable to experimental testing in future work.

RDC is not necessarily in conflict with existing theories, because RDC instantiates nothing more than the very general suppositions of a resource-rational approach to cognitive modeling. RDC states simply that the cognitive cost of taking certain actions is determined by a trade-off of minimizing loss while also minimizing information-processing costs, which can be measured using mutual information. Any model that involves minimizing loss under computational constraints will be subject to the bounds described by rate--distortion theory.

\section{General discussion}

I have shown that the rate--distortion theory of control can naturally account for similarity-based interference in general, and that it offers a strong model of picture--word/Stroop interference effects. Now I turn to the interpretation of the model and how it relates to word production more generally.

\subsection{Interpretation of computation and decision cost} 

I used two notions of cost: computation cost and decision cost, where computation cost is the cost term that is contained in the control objective, and decision cost is the cost of selecting a single action given a probabilistic policy. As a summary, semantic similarity-based interference emerged in the decision cost, while computation cost predicted general difficulty for the less-probable goal in context (naming as opposed to reading). 

It should be emphasized that decision cost, as I have presented it, is simply the cost associated with sampling a single deterministic action from a probability distribution over actions.
In the interference model, we have two sources of input information, the behavioral goal $g$ and the perceptual state $s$, and the computation cost reflects the difficulty of computing a distribution $q(a|g,s)$ over output actions given these input information sources. 
The decision cost, in contrast, does not reflect any further information processing involving $g$ or $s$ directly, although it may reflect a cost of sampling from $q(a|g,s)$ at a higher or lower temperature (such that the actual probability of an action may not correspond directly to $q(a|g,s)$).
It may be that other linking functions (other than the logarithm used here) provide a better connection between $q(a|g,s)$ and empirically observable response times, for example functions based on drift-diffusion models \citep{ratcliff2008diffusion}.

\subsection{Further word production phenomena}

I intend to advance RDC, or an extension of it, as a model of word production in general. I have presented its application to PWI and Stroop interference because these are well-known and challenging phenomena to model. However, there are many other language production phenomena on which an RDC model has yet to be tested.




One such phenomenon is phonological facilitation. The PWI task famously exhibits phonological facilitation, meaning that naming time is sped up when the distractor word is \emph{phonologically} similar to the target word \citep{meyer1991phonological}. In the simple simulations presented here, the RDC does not predict this kind of facilitation. However, it can when the control objective is specified in more detail, as I sketch below.

\begin{figure}
    \centering
    \begin{tikzpicture}
        \node (S) at (0,0) {$S$};
        \node (G) at (0,2) {$G$};
        \node (A) at (2,1) {$A$};
        \draw[->] (G) -- (A);
        \draw[->] (S) -- (A);
    \end{tikzpicture}
    \hspace{3cm}
    \begin{tikzpicture}
        \node (S) at (0,0) {$S$};
        \node (G) at (0,2) {$G$};
        \node (A1) at (2,0) {$A_1$};
        \node (A2) at (2,2) {$A_2$};
        \draw[->] (G) -- (A1);
        \draw[->] (S) -- (A1);
        \draw[->] (G) -- (A2);
        \draw[->] (S) -- (A2);        
    \end{tikzpicture}    
    \caption{\textbf{Left}, schematic of the policy in Eq.~\ref{eq:synergy-penalty}, where the behavioral goal $G$ and the perceptual state $S$ jointly determine a single atomic action $A$. \textbf{Right}, a policy where the behavioral $G$ and the perceptual state $S$ determine two actions $A_1$ and $A_2$ to be performed by different actuators.}
    \label{fig:multivariate-policy}
\end{figure}
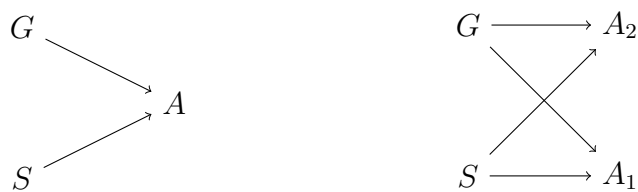

Imagine that the goal of the policy is not to output a single atomic output action, but rather to output a large number of actions. For example, one can imagine that the policy must output instructions to a large number of actuators. This kind of policy is illustrated in Figure~\ref{fig:multivariate-policy}. Equivalently, the output of the policy is a vector $\mathbf{a} = [a_1, a_2, \dots, a_n]$ of actions to be performed by $n$ different actuators.

Given this kind of policy, we can define a `phonological' similarity metric among actions $\mathbf{a}_1$ and $\mathbf{a}_2$ in terms of how many elements overlap between $\mathbf{a}_1$ and $\mathbf{a}_2$. For each overlapping element, we will have a facilitation effect, and for each non-overlapping element, we will have an interference effect. The result is overall facilitation when the target action and the distractor have more overlapping elements.

There are other extensions of RDC and other mechanisms that could give rise to facilitation-like effects. Families of more elaborate RDC policies have been explored in simulations by \citet{genewein2015bounded}.

\subsection{Conclusion}

This work has extended the reach of information-theoretic models of language processing. Although information-theoretic models have seen broad success in the study of language comprehension \citep{hale2001probabilistic,moscosodelpradomartin2004putting,levy2008expectation,hale2018finding,futrell2020lossy} and the emergence of linguistic structure \citep{zaslavsky2018efficient,hahn2020universals}, they have not yet seen much application to language production. This work has taken the first steps toward remedying this gap using the rate--distortion theory of control. 


\section*{Acknowledgments}

I would like to thank Seth Frey, Lisa Pearl, and Greg Scontras for discussion.

\bibliography{everything}
\bibliographystyle{acl_natbib}

\end{document}